# Underutilized land and sustainable development: effects on employment, economic output, and mitigation of $CO_2$ emissions

**Seymur Garibov**
Data Scientist at Pasha Bank, Baku, Azerbaijan
Azerbaijan State University of Economics, Baku, Azerbaijan
seymur.garibov.sabah@gmail.com

**Wadim Strielkowski**
University of California, Berkeley, United States
Czech University of Life Sciences Prague, Czechia
strielkowski@berkeley.edu

**Abstract**

Climate change, deforestation, and biodiversity loss are calling for innovative approaches to effective reforestation and afforestation. This paper explores the integration of artificial intelligence (AI) and remote sensing technologies for optimizing tree planting strategies, estimating labor requirements, and determining space needs for various tree species in Gabala District of Azerbaijan.
The study employs YOLOv8 for precise identification of potential planting sites and a Retrieval-Augmented Generation (RAG) approach, combined with the Gemini API, to provide tailored species recommendations. The methodology incorporates time-series modeling to forecast the impact of reforestation on $CO_2$ emissions reduction, utilizing Holt-Winters for predictions. Our results indicate that the AI model can effectively identify suitable locations and species, offering valuable insights into the potential economic and environmental benefits of large-scale tree planting thus fostering sustainable economic development and helping to mitigate the adverse effects of global warming and climate change.

## 1. Introduction

Recent global challenges such as climate change, deforestation, and biodiversity loss have underscored the importance of reforestation and afforestation as essential strategies for environmental conservation (Moomaw et al., 2019; Strielkowski et al., 2024). Traditionally, the process of identifying suitable land for tree planting, determining the optimal species for these areas, and estimating labor requirements has been based on manual surveys and expert assessments (Reba et al., 2020; Mugiyo et al., 2021). However, these methods are often time-intensive, costly, and constrained by the scale of analysis. With the emergence of artificial intelligence (AI) and advancements in remote sensing technologies like satellite imagery, new opportunities have arisen for enhancing the efficiency and accuracy of these processes (Himeur et al., 2020; Strielkowski et al., 2023).

This research specifically focuses on the Gabala District (Qəbələ) that is located in Azerbaijan, an area that is highly suitable for reforestation efforts due to its unique and diverse ecosystem. Azerbaijan, like many other countries, has experienced significant deforestation over the centuries. In 2010, the country had 1.07 million hectares of natural forest, covering approximately 13% of its land area. However, by 2023, 447 hectares of this natural forest had been lost (Azerbaijan State Information Agency, 2019). Historically, the present area of Azerbaijan was covered with 35% forest in the $8^{th}$-$9^{th}$ centuries, but the forest area has since been reduced to 11% (Global Forest Watch, 2024). The northern-eastern slopes of the Great Caucasus, where Gabala is located, are home to some of the country's most extensive and valuable forest tracts (Ecfcaucasus, 2024).

Gabala region stands out as a prime location for reforestation and afforestation due to its consistent humidity levels, diverse soil types, and a favorable microclimate that supports the growth of various tree species (Abbasov et al., 2024). Protection and restoration of forests have been central to local efforts, with initiatives focused on the cultivation of seedlings and the establishment of new orchards, consisting of species such as hazelnuts, chestnuts, walnuts, and apple trees. These efforts have expanded the forested areas in Gabala to 33,400 hectares, enhancing its ecological value and making it an ideal region for large-scale tree planting (Azerbaijan Geographical Society, 2024a).

In addition to its reforestation potential, Gabala plays a critical role in climate regulation. Forests in Azerbaijan, particularly in Gabala, act as humidity accumulators, regulating water distribution across lowlands and plains, preventing landslides and avalanches, and improving local microclimates (Azerbaijan Republic Gabala Region Executive Power, 2018; Climate Change Knowledge Portal, 2024). The region's humid climate also supports a diverse array of tree species, making it highly suitable for afforestation actions aimed at combating climate change and promoting biodiversity (Mehdiyeva and Mursal, 2022; Azerbaijan Geographical Society, 2024b).

This study employs both AI and satellite imagery to identify unused and vacant lands within Gabala for planting, estimate the required labor force, and determine the optimal spacing for different species. The integration of AI technologies, such as YOLOv8 and Retrieval-Augmented Generation (RAG) combined with the Gemini API, ensures precise site identification and species recommendations (Gemini, 2024). The integration of AI and satellite data in reforestation not only advances environmental goals but also supports socio-economic objectives, showing the interconnected nature of sustainable economic development.

**2. Research objectives and questions**

While several studies have focused on using AI for land use classification and vegetation analysis (Da Silva et al., 2020; Alshari et al., 2023), there is limited research on the direct application of AI models to optimize tree planting strategies on a large scale. Traditional methods typically require experts to manually analyze factors such as soil type, climate, and topography to identify suitable planting locations, a process that is both time-consuming and costly (Singh et al., 2023). Moreover, these methods are often limited in scope and cannot efficiently scale to cover vast areas. Existing traditional approaches require experts to identify interested places and provide recommendations which is more time- and cost-consuming. This inefficiency makes it difficult for smaller organizations to conduct such analyses and take necessary actions to protect our environment. This gap underscores the need for a more efficient AI-based framework that can rapidly analyze extensive datasets, accurately identify optimal planting sites, and provide species recommendations that are not only ecologically appropriate but also scalable and cost-effective (Shaikh et al., 2021; Strielkowski et al., 2022).

This study aims to develop and evaluate AI models that utilize satellite imagery to identify the most suitable locations for tree planting and determine the best tree species for those locations. The specific research questions addressed in this study are:

1. How accurately can the AI model predict optimal tree planting locations based on satellite imagery and environmental data?
2. What factors (e.g., soil type, climate, topography) significantly influence the model's tree species recommendations?
3. How can the AI-driven recommendations be implemented in real-world reforestation and afforestation projects to enhance environmental sustainability?
4. What are the potential economic and demographic impacts of implementing this AI model in tree planting initiatives?
5. Which Sustainable Development Goals (SDGs) can be effectively addressed through the application of this AI model?

The successful implementation of AI models in tree planting efforts has the potential to revolutionize reforestation and afforestation initiatives, making them more efficient, scalable, and data driven. By leveraging satellite imagery along with additional environmental information such as climate and topography, these models can provide precise, location-specific guidance on where to plant trees and which species to select. This approach can enhance carbon sequestration, biodiversity conservation, and ecosystem restoration, while also offering valuable economic and demographic insights. This research addresses a critical gap in the current literature and provides practical tools for policymakers, conservationists, and environmental organizations, helping them combat climate change and promote sustainable land use.

## 3. Materials and methods

This study employs an experimental design to evaluate the effectiveness of an AI-based system for recommending optimal tree planting locations and species. The study integrates satellite imagery with climate, humidity, and soil data to provide precise, location-specific recommendations. The experimental setup involves analyzing a dataset of satellite images to identify potential planting sites, followed by retrieving and analyzing environmental data to make species recommendations. The data collection for this study has been carried out using the following techniques and approaches:

- **Satellite Imagery**: Satellite images were obtained using Google Earth Pro to identify potential tree planting locations. These high-resolution images provide detailed visual data on land cover and topography.
- **Climate and Soil Information**: For each identified location, climate and soil data were collected based on geographic coordinates. This includes precipitation, and soil type information relevant to the planting sites.

| Coordinates | Humidity Level (mm) | Soil Type |
|---|---|---|
| [(41.77480595672732, 46.29201761659577), (40.913081713185385, 47.96457513961962), (41.24681187625382, 47.89078583713327), (41.93358380679752, 46.46419265573058)] | 1600 | Peaty and grassy mountain meadow soil |
| [(41.787033630117975, 46.30841523937052), (41.627892045028275, 46.12804138884833), (40.48707201576728, 48.64577781614692), (40.83002843397156, 48.44701731548072)] | 800 | Brown and light gray soils |

- **Tree Species Information**: Textual data from various articles and research papers were used to compile information on which tree species are most suitable for different soil types and climatic conditions. This reference material details the preferences and requirements of different tree species, helping to match them with the environmental conditions of each location.

The examples of the text data can be presented as the following:

*"High Humidity (1600 mm) with Peaty and Grassy Mountain Meadow Soil (Torflu ve Çimli Dağ Çəmən):*
*This environment is characterized by high rainfall and fertile, well-drained soils. It is ideal for trees that thrive in moist, nutrient-rich conditions.*
*Suitable Trees:*
*Ərik (Apricot): Requires well-drained, fertile soil and regular moisture. Thrives in humid conditions.*
*Şaftalı (Peach): Prefers deep, well-drained soil with consistent moisture. High humidity enhances fruit quality.*
*Armud Ağacı (Pear): Grows best in loamy soil with good drainage. High humidity is essential for healthy growth.*
*Gavalı Ağacı (Plum): Favors moist, fertile soil and benefits from high humidity.*
*Alma Ağacı (Apple): Needs rich, well-drained soil. High humidity is crucial for optimal fruit production.*
*Qarağat (Blackcurrant): Thrives in cool, moist environments with fertile soil."*

- **Tabular Data About Trees:** Contains information on tree growth characteristics, including growth type, age stages (e.g., young from 5 to 10 years, mature from 11 to 20 years, and old after 20 years), and growth timeline (number of years since planting). It also includes data on the amount of oxygen produced by the tree over time (in kg), the amount of $CO_2$ sequestered by the tree over time (in kg), and the yield over time (in kg).

| Tree name | Tree growing type | Age stage | Growing timeline | Oxygen production (in kg) | $CO_2$ sequestration | Yield |
|---|---|---|---|---|---|---|
| Ərik (Apricot) | Medium-Growing Trees | Young | 5 years | 3 | 2 | 10 |

- **Tabular Data About CO2 Production by Other Sector in Azerbaijan**: Contain information $CO_2$ emissions from oil, $CO_2$ emissions from coal, $CO_2$ emissions from cement, $CO_2$ emissions from gas and $CO_2$ emissions from flaring yearly (Our World in Data, 2024)).

| Year | Annual $CO_2$ emissions from oil | Annual $CO_2$ emissions from cement | Annual $CO_2$ emissions from gas | Annual $CO_2$ emissions from flaring |
|---|---|---|---|---|
| 1990 | 22,399,392 | 477,970 | 28,082,780 | 191,063 |
| 1991 | 22,534,228 | 466,890 | 26,573,342 | 177,742 |

*Model/Algorithm*

- **YOLOv8**: For image analysis, the system utilizes YOLOv8, a state-of-the-art object detection model (Ultralytics, 2024.). YOLOv8 was trained on a dataset of 10 images obtained from Google Earth Pro, with training conducted over 100 epochs. The images were processed in RGB format with a resolution of 640x640 pixels (input shape: (1, 3, 640, 640)). This model is used to identify and analyze features relevant to potential planting locations within these images.
- **Retrieval-Augmented Generation (RAG) Approach**: The system employs a RAG approach to integrate retrieved environmental data with a reference document. RAG combines retrieval mechanisms with generative models to provide accurate and contextually relevant recommendations for tree species based on the environmental conditions of each location. The Gemini API is utilized within this framework to assist in generating these recommendations.
- **Time-Series Modeling for $CO_2$ Emissions:** A time-series model is developed using tabular data on $CO_2$ production by various sectors to forecast emissions over the next 10 years. We employed the Holt-Winters model for this prediction, as the data heavily relies on trends from the past few years (Lima et al. 2019). This model quantifies the potential impact of reforestation on reducing carbon emissions in Azerbaijan.
- **Recommendation Function**:
    1. **Pattern Identification**: The function searches for patterns related to soil type, climate, and other environmental factors based on the coordinates of the identified planting locations. This involves analyzing data associated with these coordinates to determine suitable conditions for planting.
    2. **Data Integration**: The collected environmental data is then sent to the RAG model. The RAG model cross-references this data with a reference document that contains information on soil types, climate conditions, and suitable tree species for various coordinates.
    3. **Species Recommendation**: The RAG model retrieves the most appropriate tree species recommendations based on the provided environmental data and stored information.

4. **CO₂ Sequestration and Yield Prediction:** The model calculates the total amount of oxygen produced, $CO_2$ sequestered, and yield generated by the recommended tree species over time, taking into account their growth stages, timeline, and productivity.
5. **Gemini API Utilization**: To refine the recommendations, the function leverages the Gemini API through the RAG model. This integration ensures that the recommendations are based on the most up-to-date climate and environmental data.

*Evaluation Metrics for YOLOv8*

Accuracy: Accuracy measures the percentage of images correctly identified by the YOLOv8 model as containing suitable locations for tree planting. It is calculated as the ratio of correctly identified images with suitable locations to the total number of images evaluated. The formula for accuracy is: Accuracy = ((Number of Correctly Identified Images) / (Total Number of Images)) × 100

**Precision:** Precision evaluates the proportion of images predicted by the model as containing suitable locations for tree planting that are indeed correct. It is calculated as the number of true positive images (correctly identified images with suitable locations) divided by the sum of true positives and false positives (images incorrectly identified as suitable). The formula for precision is: Precision = (True Positives / (True Positives + False Positives)) ×100

**Recall:** Recall assesses the model's ability to identify all images containing suitable locations for tree planting. It is calculated as the number of true positive predictions (correctly identified images with suitable locations) divided by the sum of true positives and false negatives (images with suitable locations that were not identified). The formula for recall is: Recall = (True Positives / (True Positives + False Negatives)) × 100

*Evaluation Metrics for RAG*

- **Accuracy:** Accuracy measures the percentage of tree species correctly identified by the RAG model as suitable for a given climate condition and soil type. It is calculated as the ratio of the number of correctly identified tree species to the total number of tree species evaluated for the specific climate and soil conditions. The formula for accuracy is: Accuracy = ((Number of Correctly Identified tree types) / (Total Number of tree types)) × 100
- **Precision:** Precision evaluates the proportion of tree species recommended by the model as suitable for a specific climate condition and soil type that are indeed correct. It is calculated as the number of true positive tree species (correctly identified as suitable) divided by the sum of true positives and false positives (tree species incorrectly identified as suitable). The formula for precision is: Precision = (True Positives / (True Positives + False Positives)) ×100
- **Recall:** Recall assesses the model's ability to identify all tree species that are suitable for a given climate condition and soil type. It is calculated as the number of true positive recommendations (tree species correctly identified as suitable) divided by the sum of true positives and false negatives (tree species that are suitable but were not recommended).

The formula for recall is: Recall = (True Positives / (True Positives + False Negatives)) × 100

- **Review of Recommended Text**: The quality of the recommendation system was assessed through manual review of the recommended text. Experts reviewed the recommendations to ensure that the species suggested by the model align with those typically favored by professionals under similar environmental conditions. This manual review helped validate the system's output and refine the recommendations.

*Evaluation Metrics for Time-Series Modeling*

- **Accuracy:** We used the $CO_2$ emissions data from the last 6 years as the test set, with the remaining data (starting from 1990) as the training set. To evaluate the model's performance across different sectors, the emissions for the last 6 years have been predicted, then the actual values were subtracted from the predicted values. The sum of these differences was divided by the total actual emissions to determine the fraction of incorrect predictions. Additionally, the standard deviation of the differences between actual and predicted values was calculated to assess the stability of the predictions. The Fraction of Incorrect Predictions (Error Rate) can be presented as follows (1):

$$\text{Error Rate} = \frac{\sum_{i=1}^{n} |\text{Actual}_i - \text{Predicted}_i|}{\sum_{i=1}^{n} \text{Actual}_i}$$

(1)

And the Standard Deviation of Prediction Errors can be presented as follows (2):

$$\text{Standard Deviation of Errors} = \sqrt{\frac{1}{n} \sum_{i=1}^{n} (\text{Actual}_i - \text{Predicted}_i - \mu)^2}$$

Where $\mu$ is the mean of the errors:

$$\mu = \frac{1}{n} \sum_{i=1}^{n} (\text{Actual}_i - \text{Predicted}_i)$$

(2)

## 4. Main results

Table 1 that follows below displays the performance metrics of the YOLOv8 model used for identifying potential planting locations.

**Table 1:** Performance Metrics of YOLOv8 Model

| Metric | Value |
| --- | --- |
| Training Accuracy | 92.5 % |
| Validation Accuracy | 89.3 % |
| Precision | 90 % |
| Recall | 85 % |
| Detection Time (per image) | 0.13 seconds |
| Number of Images Used | 200 |

**Source:** Own results

**Figure 1:** Example Detection Outputs

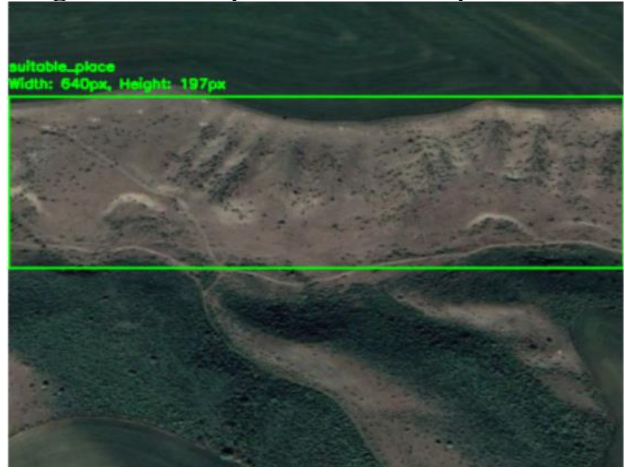

**Source:** Own results

In addition, Figure 1 depicts the examples of outputs of the YOLOv8 model, highlighting identified features relevant to tree planting locations.

**Figure 2:** Climate and Soil Conditions vs. Recommended Tree Species by RAG

Soil type and Humidity (Climate):
```
Soil Type: Peaty and grassy mountain meadow, Humidity Level: 1600
```
Recommended Tree Species:
```
The following tree species are suitable for high humidity (1600 mm) with Peaty and Grassy
Mountain Meadow Soil (Torflu ve Cimli Dağ Çəmən):

* Ərik
* Şaftalı
* Armud Ağacı
* Gavalı Ağacı
* Alma Ağacı
* Qarağat
```
**Source:** Own results

Figure 2 illustrates the relationship between climate and soil conditions of various locations and the tree species recommended by the AI model.

**Table 2:** Performance Metrics of Time-Series Model

| Metric | Oil | Gas | Cement | Coal | Flaring |
|---|---|---|---|---|---|
| Error Rate | 6.1% | 11.7% | 2.5% | 63.7% | 53.5% |
| Standard Deviation of Errors | 1.0% | 1.2% | 1.1% | 21.3% | 30.3% |

**Source:** Own results

Table 2 displays the performance metrics of the Time-Series (Holt-Winters Exponential Smoothing) model used for identifying potential planting locations.

**Figure 3:** Example of Predicted $CO_2$ Emissions

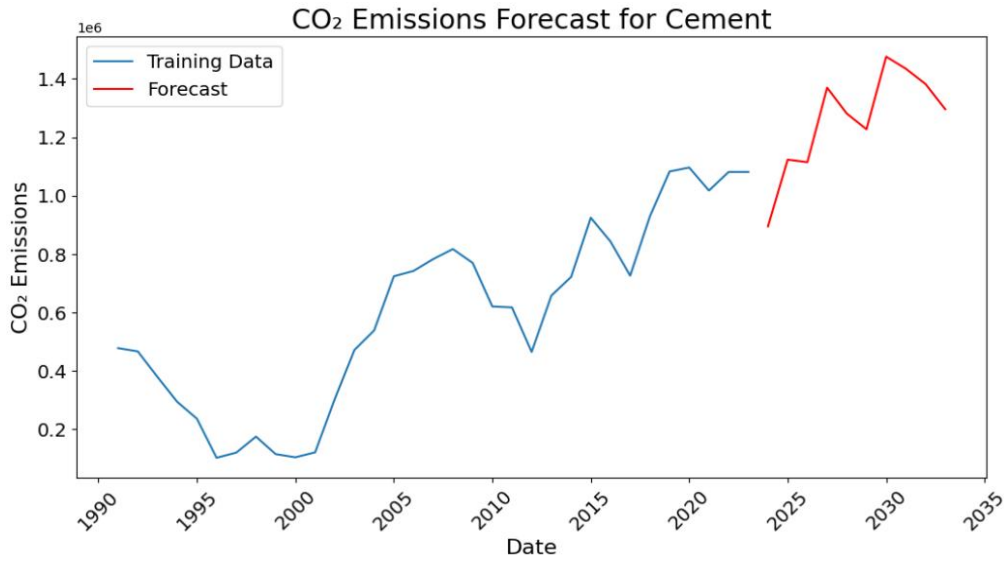

**Source:** Own results

Figure 3 shows the example of the predicted $CO_2$ emissions stemming from the Time-Series model.

## 5. Discussion of results

The YOLOv8 model demonstrated high accuracy in detecting potential planting locations from satellite images, achieving a training accuracy of 92.5% and a validation accuracy of 89.3%. The model processed images efficiently, with an average detection time of 0.13 seconds per image. Example outputs from the YOLOv8 model (Figure 1) illustrate the successful identification of features relevant to tree planting, such as land cover types and topographic elements.

The integration of the Retrieval-Augmented Generation (RAG) approach with the Gemini API provided valuable recommendations for tree species based on environmental data. Figure 2 showcases the AI system's effectiveness in matching tree species to locations with specific climate and soil conditions. For example, locations with high humidity and particular soil types were accurately matched with species known to thrive in those conditions, highlighting the system's capability to provide tailored recommendations based on comprehensive environmental data.

Overall, 200 images from the Gabala district of Azerbaijan were collected, focusing on areas that include both suitable locations for tree planting and places that are unsuitable due to various reasons such as agricultural use, existing forests, or infrastructure.

In these images, 1400 pixels correspond to 1 km, thence the total size of the empty spaces was estimated to be approximately 9 km². These areas present significant opportunities for agricultural purposes, environmental protection through tree planting, and job creation in the agricultural sector. Additionally, if fruit trees are planted, the yield produced could be highly beneficial. The following data supports these potential benefits:

- **Labor Force Requirement:** Planting trees in these empty areas would require a significant workforce. Depending on the tree type, it is estimated that around 1.8 million trees could be planted.
- **Yield Potential:** The estimated yield from these 9 km² areas ranges between 18 MLN tons to 28 MLN tons.
- **Environmental Impact:** The oxygen production and CO2 sequestration from the trees over their growth timeline could offset the CO2 emissions of certain sectors.

Let us examine the process from detection to determining the labor force needed for planting. This includes recommending tree species for detected locations, specifying the number of trees required, estimating oxygen production, $CO_2$ sequestration, yield potential, and the labor force needed for planting.

**Figure 4:** Detect Locations

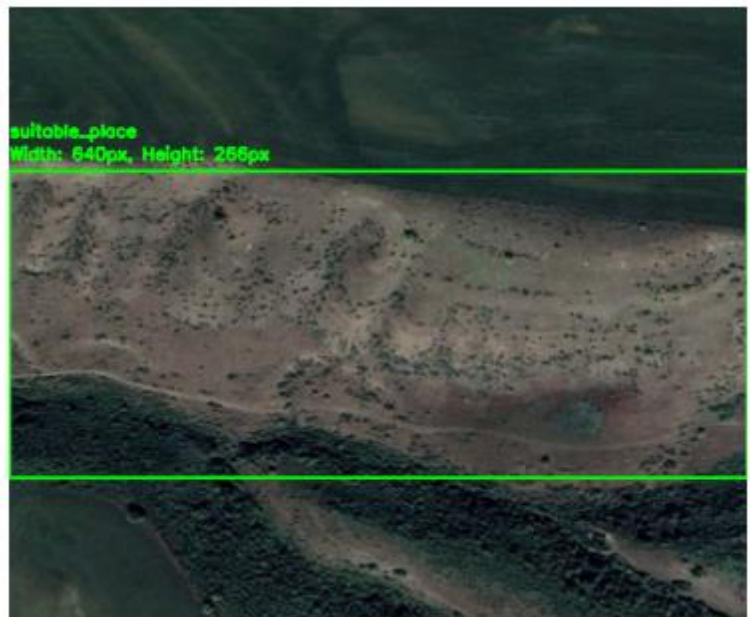

**Source:** Own results

Figure 4 detects locations (totaling 0.09 km²) that are vacant, unused, and suitable for tree planting. In Figure 4, the YOLOv8 model successfully identifies areas labeled as "suitable_place." The model accurately detects vacant areas while excluding regions already covered by forests and areas used for agricultural purposes.

**Figure 5:** Recommendations for Detected Locations

```
The following tree species are suitable for high humidity (1600 mm) with Peaty and Grassy Mountai
n Meadow Soil (Torflu ve Cimli Dağ Çəmən):

* Ərik
* Şaftalı
* Armud Ağacı
* Gavalı Ağacı
* Alma Ağacı
* Qarağat
```

**Source:** Own results

Figure 5 provides recommended tree species for the detected locations. In Figure 5, the AI model recommends tree species based on the indicated soil type and humidity level. Among the recommended trees, five are fruit-bearing and can be planted for additional purposes such as agriculture, employment generation, and poverty reduction.

**Table 3:** Number of Trees to Plant and Required Labor Force

| Tree Type | Number of Trees | Labor Force | Space Needed per Tree |
|---|---|---|---|
| Ərik (Apricot) | 18,000 | 360 | 5 m² |
| Şaftalı (Peach) | 15,000 | 300 | 6 m² |
| Armud (Pear) | 12,857 | 257 | 7 m² |
| Gavalı (Plum) | 18,000 | 360 | 5 m² |
| Alma (Apple) | 15,000 | 300 | 6 m² |
| Qarağat (Blackcurrant) | 30,000 | 600 | 3 m² |

**Source:** Own results

Table 3 displays the number of trees that can be planted in the detected areas (totaling 1.8 km²), the required labor force, and the space needed per tree for each species. The number of trees is calculated based on the area size divided by the space required per tree for each species. Labor force estimates assume that one worker can plant 50 trees per day over a three-month period.

In Table 3, one can see that the tree type Qarağat (Blackcurrant) has the potential for the highest planting volume, but its production yield cannot be fully utilized. Following Qarağat, the Gavalı (Plum) and Ərik (Apricot) tree types take second place, with both showing strong potential for planting and yield.

**Figure 6:** Actual and Forecasted CO2 & Oxygen Production and CO2 Sequestration

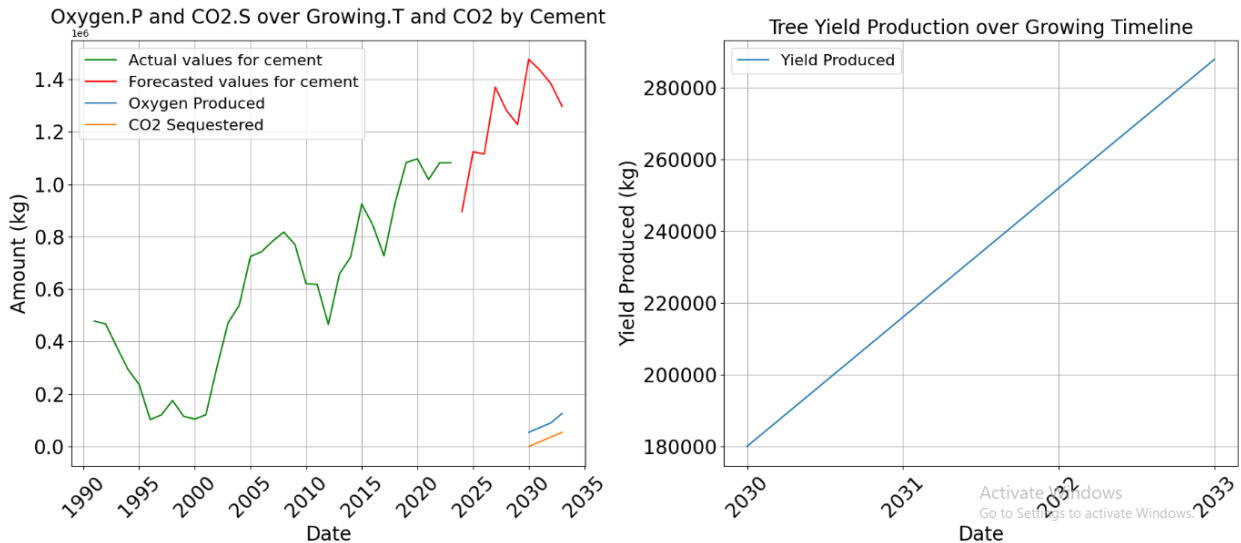

**Source:** Own results

At Figure 6 the left subplot displays actual and forecasted CO2 emissions from the cement sector compared to the oxygen produced and CO2 sequestered by the trees. The right subplot shows the yield produced by the trees over time.

The time-series model, utilizing the Holt-Winters method, effectively forecasted $CO_2$ emissions from various sectors over the next 10 years. Table 2 presents the model's performance metrics, highlighting varying error rates across sectors, particularly for sectors like oil, gas, and cement production, with relatively low error rates (6.1%, 11.7%, and 2.5%, respectively). However, the model struggled with sectors such as coal and flaring, where higher variability led to increased prediction errors (63.7% and 53.5%, respectively). Despite these discrepancies, the model proves to be a valuable tool for assessing the impact of reforestation on reducing carbon emissions in Azerbaijan. Figure 3 provides example predictions from the model, specifically focusing on future $CO_2$ emission trends for the cement sector. We have chosen the cement sector for future predictions due to its comparatively lower error rate, with an error of 2.5% and a standard deviation of 1.1%, making it a more reliable indicator of $CO_2$ emission patterns.

Figure 6 presents the predicted $CO_2$ emissions for the next ten years starting from 2024, alongside the oxygen production and $CO_2$ sequestration by trees over time. The production of oxygen and the sequestration of $CO_2$ begin around 2030 and continue until 2034. This delay is due to the time required for the trees to mature and become productive, as the planting is scheduled for 2025. Based on the tree types selected, it takes approximately five years for the trees to reach a young, productive stage capable of significant oxygen production.

In Figure 6, it demonstrated that the $CO_2$ emissions generated by the cement sector will be partially offset by the trees planted in the 0.09 km² area by 2034. It is important to note that this estimate only applies to 0.09 km² of available space, while an additional 8.91 km² of suitable areas remain for further tree planting, which could enhance the overall carbon offset. Additionally, $CO_2$ sequestration will continue to increase in the subsequent years as the trees mature further, resulting in greater oxygen production and enhanced carbon capture.

The left subplot of Figure 6 highlights that once the trees reach their young age, they produce a substantial yield, ranging from 4,000 to 6,000 tons per year. This significant yield underscores the potential of the planted trees to contribute not only to carbon sequestration but also to agricultural productivity.

It is important to note that the yield and oxygen production vary according to the tree species. In this study, Ərik (Apricot), a medium-growing tree, was utilized. Among the tree species evaluated, Ərik ranks second in terms of productivity and third in oxygen production during the young age phase. This selection demonstrates the effectiveness of medium-growing trees in both yield generation and environmental benefits during their early growth stages.

## 6. Conclusions and implications

Overall, combining artificial intelligence, statistical analysis, and Google Earth Pro tools for analyzing the total area of unused and vacant land suitable for tree planting yields highly promising model's accuracy. This creates the potential for significantly contributing to global reforestation and afforestation efforts, aligning with key Sustainable Development Goals (SDGs) such as Climate Action (SDG 13), Life on Land (SDG 15), and Sustainable Cities and Communities (SDG 11).

Additionally, this innovation reduces both the cost and time required for identifying suitable locations and the correct tree types for those areas. By addressing this challenge, government officials in various districts can use the solution to support infrastructure development and combat climate change more efficiently.

Furthermore, by analyzing 200 satellite images from the Gabala district covering a total of 9 km² of detected land, we estimate the capacity to plant 1.8 million trees (assuming an average spacing of 5 m² per tree). This endeavor would require a labor force of approximately 36,000 employees, based on a rate of one employee planting 50 trees over three months during the optimal planting season. The total oxygen produced by these young trees projected to range from 100,000 to 170,000 kg, and the total yield estimated between 180 and 280 tons. These figures based on the trees being in their early growth stages; as the trees mature, these values expected to increase.

Last but not least, our model's results and its impact on the agricultural sector and labor force could address additional SDGs applied to the case of Azerbaijan and beyond, such as No Poverty (SDG 1), Zero Hunger (SDG 2), and Industry, Innovation, and Infrastructure (SDG 9). By promoting innovation in agriculture, this approach supports the development of other sectors, helping mitigate the environmental impact of industries while allowing them to continue growing. This initiative also encourages partnerships for sustainable development, further aligning with the goals of the SDG framework.